# A PROGRAM FOR THE GEOMETRIC CLASSIFICATION OF PARTICLES AND RELATIVISTIC INTERACTIONS

**Gustavo González-Martín**

Departamento de Física, Universidad Simón Bolívar,

Apartado 89000, Caracas 1080-A, Venezuela.

Web page at http:\\prof.usb.ve\ggonzalm\

Geometric relativistic interactions in a new geometric unified theory are classified using the dynamic holonomy groups of the connection. Physical meaning may be given to these interactions if the frame excitations represent particles. These excitations have algebraic and topological quantum numbers. The proton, electron and neutrino may be associated to the frame excitations of the three dynamical holonomy subgroups. In particular, the proton excitation has a dual mathematical structure of a triplet of subexcitations. Hadronic, leptonic and gravitational interactions correspond to the same subgroups. The background geometry determines non trivial fiber bundles where excitations live, introducing topological quantum numbers that classify families of excitations. From these particles, the only stable ones, it may be possible, as suggested by Barut, to build the rest of the particles. The combinations of the three fundamental excitations display $SU(3)\otimes SU(2)\otimes U(1)$ symmetries.

# 1. Introduction.

The group of the geometric space-time structure of special relativity is fundamental to the field theories of elementary particles, which are representations of this group. In contrast the geometry of general relativity has not played such a fundamental role. Nevertheless, our geometric relativistic unified theory of gravitation and electromagnetism may have non trivial applications for particle theory [1,2]. Here we discuss this question using groupal and geometric features and ideas, avoiding unnecessary aspects that may present obstacles in understanding the physical implications of the geometry for physical particles. The study of groups that act on the geometric structures of a physical theory and their related symmetric spaces may display essential physical features without actually solving the equations of the theory. We hope to show the relation of this particular new approach in the treatment of elementary particles and their interactions, a geometric group realization as fiber bundle sections, with the standard model [3,4].

It is well known that the holonomy groups of a connection in a fiber bundle may be used to classify its possible connections. It is interesting to use this method to gain insight on the types of physical interactions represented by the connection in the unified geometric theory.

In QFT certain particles are given masses by a Higgs mechanism which relies on certain symmetries that a vacuum solution may possess. This appears to assign to the vacuum a not completely passive role. We may try to assign to the vacuum an even more active role. This is accomplished by recognizing that the particle vacuum is a geometric space with physical meaning, related to a unified non linear geometric theory of physics. We consider an approximation, to the geometric non linear theory, where the microscopic physical objets are realized as linear geometric excitations around a non linear substratum geometric space. This is consistent with the QFT interpretation of particles as vacuum excitations. We interpret the excitation as particles and the substratum as the particle vacuum.

With this definition, a particle is acted upon by the substratum and is never really free except in absolute empty space (zero substratum curvature). The substratum space carries the universal inertial properties. A free particle is an idealization. In a fundamental level, if we accept physics in the substratum we are assuming, in part, a (Parmenides) holistic principle which should be consistent with the ideas of Mach [5] and Einstein [6] that assign fundamental importance of far-away matter in determining the inertial properties of local matter. On the contrary, if we assume only particles we are assuming a (Democritus) atomistic principle. This implies there is no physics in the substratum.

Restriction to holonomy subgroups has also implications for the equations of motion of matter, since within the theory particles are represented by frame excitations (sections of a bundle). We may expect that the geometric association of frames to holonomy subgroups may naturally classify particles and associate them to interactions. Hopes to accomplish this objectives are supported on results of recent previous work [7], [8], [9] which we shall summarize as follows.

Constants associated to the substratum connection appear as constant mass parameters in the excitation equations and play a fundamental role. As in General Relativity, the covariant equations may be referred to frames (coordinates) that must be related to observers determined by the physical experiment in question. Otherwise theoretical results remain indeterminate. The freedom to choose an arbitrary reference frame by a group transformation generates a class of equivalent solutions, linking *relativistic* interactions, represented by one particular reference frame $e_r$.

Any excitation must be associated to a definite substratum. An arbitrary observation of an excitation property depends on both the excitation and the substratum, but the physical observer must be the same for both excitation and substratum. We may use the freedom to select the reference frame, to refer the excitation to the physical frame defined by its own substratum, which satisfies the nonlinear equation,

$$D^{*}\Omega_{b} = 4\pi\alpha\, {}^{*}\!J_{b} \quad . \qquad (1.1)$$

Then the substratum is referred to itself and the substratum matter frame $e_b$, referred to $e_r$ becomes the group identity $I$. Actually this generalizes comoving coordinates (coordinates adapted to dust matter geodesics) [10]. We adopt coordinates adapted to local substratum matter frames (the only non arbitrary frame is itself). Free matter shows no self acceleration, self action. In its own reference frame these effects actually disappear. Only constant self energy terms, determined by the nonlinearity of the substratum, make sense and should be the origin of the constant mass parameter.

At a small distance $\lambda$, characteristic of free excitations, the connection and frame of the substratum appears symmetric. Mathematically we should say that the substratum is a locally symmetric space [11 ] or hyperbolic manifold [12 ]. We recognize the necessary condition that locally the substratum be a bundle that locally admits a maximal set of Killing vectors of the space time symmetry with null connection Lie derivative [13 ]. This means that there are space time Killing coordinates such that the connection is constant non zero in the region of particle interest. (A flat connection is too strong assumption).

It is clear from the definition of excitation that a free particle (excitation) is a representation of the structure group of the theory and consequently an algebraic element. A representation (and therefore a particle) is characterized by the eigenvalues of the Casimir operators. The state of a representation (particle) are characterized by the eigenvalues of Cartan subspace basis operators. This provides a set of algebraic quantum numbers to the excitation. Of course, we must somehow choose the respective representations associated to these particles. It has been indicated that the physical particles are representations of holonomy subgroups of SL(4,R) induced from the subgroup SL(2,C), realized as functions on the coset spaces. In fact, it has been shown that new electromagnetic consequences of this theory leads to quantization of electric charge and magnetic flux [14 ], providing a plausible explanation for the fractional quantum Hall effect.

One important issue is: How do we calculate mass in a consistent manner? Mass arises from constants with inverse length dimension corresponding to a symmetric substratum solution of the non linear field equation. These constants appear in the linear excitation equations for a fermionic frame fluctuation $\delta e$ and a bosonic connection fluctuation $W{=}\delta\Gamma$, which are of the general form, in the defining representation,

$$\mathcal{P}(\delta e) = \left(\kappa^{\mu}\partial_{\mu} + \cdots\right)\delta e = m\delta e \quad , \qquad (1.2)$$

$$^{*}d^{*}dW - \mu^{2}W + \ \cdots \ = 4\pi\alpha \overset{*}{j} \quad , \qquad (1.3)$$

where $m$ is the fermionic mass parameter given by

$$m = \tfrac{1}{4}\,\mathrm{tr}\left(e^{-1}\kappa^{\mu}e\Gamma_{\mu}\right) = \tfrac{1}{4}\,\mathrm{tr}\left(J^{\mu}\Gamma_{\mu}\right) \qquad (1.4)$$

and $\mu$ is the bosonic mass parameter given by,

$$\mu^{2} = \tfrac{1}{4}\,\mathrm{tr}\,\Gamma_{\nu}\Gamma^{\nu} \quad , \qquad (1.5)$$

in terms of the constants of the substratum solution.

If we consider geometric excitations on a substratum, these expressions may be expanded as a perturbation around the substratum in terms of a small parameter $\varepsilon$, characterizing the excitation, indicating that the zeroth order term, the bare mass, is given entirely by the substratum current and connection, with corrections depending on the excitation self interaction. As indicated in previous work [15 ], these corrections correspond to a geometric quantum field theory.

In this manner, the particle bare mass parameters are not determined from the linear particle equation themselves, but rather from the holistic inertial non linear substratum solution. The existence of a constant connection solution [16 ] for these asymptotic fields (free particle substratum) provides a fundamental distance, due to non linearity, and gives a mechanism to calculate mass ratios in terms of volumes of respective symmetric coset spaces. The extreme case of absolute empty space that implies zero substratum curvature self interaction, is only mathematically possible in this theory for a universe without matter, because of the nonlinearity of the substratum field equations.

Nevertheless, the association of a mass to one of these geometric distances remains arbitrary because this association actually calibrates a geometric scale of distance in terms of a physical mass scale. Since mass is inverse length, we may associate a “standard physical mass” to the “standard geometric length”. After the association we may attempt to calculate a ratio of masses. For example, if we select the mass of the proton excitation to calibrate the geometric unit, we may calculate the mass of the electron excitation. The result of this calculation, using the symmetric space quotients of holonomy groups and the constant substratum solution is the ratio [17]

$$\frac{m_G}{m_H}=\frac{V\left(K_R\right)}{V\left(C_R\right)}=6\pi^5=1836.1181\approx\frac{m_p}{m_e} \quad , \tag{1.6}$$

which essentially agrees with the experimental value and may lend credibility to this geometric theory.

## 2. Geometric Classification of the Connection.

In order to classify the relativistic interactions we look for dynamical holonomy groups *H* of the associated connection. It is clear that *H* must be a subgroup of the structure group of the theory, SL(4,R), chosen to be the simple group of automorphisms of the space time Clifford algebra.

From the geometric meaning of the algebra, the elements $\pm\kappa_a$ are associated to triads of opposite orientation. In principle, both sets with opposite signs may be used as part of the orthonormal set $\kappa_\alpha$ of the algebra. The arbitrary sign is determined by a standard relation of products of Pauli matrices, in terms of Hodge duality $\varepsilon_{ijk}$ and Clifford (complex) duality *i*. We should adhere to the same convention in the choice of sign for the $\pm\kappa_a$. The standard orientation in space time $[u_0,u_1,u_2,u_3]$ induces an orientation in the geometric algebra which should be used to define the Clifford geometric duality $\kappa_0\kappa_1\kappa_2\kappa_3$. This four dimensional duality operation denoted by $\kappa_5$, and Hodge duality should relate the matrices representing the Pauli matrices within the algebra, preserving the standard relation. Mathematically,

$$\kappa_0\kappa_1\kappa_2\kappa_3=\kappa_5\equiv i=\sigma_0\sigma_1\sigma_2\sigma_3=\kappa_0\kappa_1\kappa_0\kappa_2\kappa_0\kappa_3 , \tag{2.1}$$

$$\sigma_i=\kappa_0\kappa_i . \tag{2.2}$$

The group SL(4,R) is generated by exponentiation of the 15 traceless matrices of a base in the Clifford algebra.

For physical reasons we want that the relativistic interactions be associated to a dynamic evolution of the matter sources. The dynamical effects are produced in the theory by the action of the group, in particular accelerations should be produced by generators equivalent to the Lorentz boosts as seen by an observer. Therefore, we require that the boost generators, $\kappa_0\kappa_\alpha$, be present in a connection identifiable with a dynamical interaction as seen by the observer associated to the Minkowski subspace generated vectorially by $\kappa_\alpha$. Because of the nature of the source current,

$$J^\mu=\tilde{e}\,\kappa^{\hat{\alpha}}u^\mu_{\hat{\alpha}}e \quad , \tag{2.3}$$

which corresponds to the adjoint action of the group on the algebra, it should be clear that all generators of the form $\kappa_{[\alpha}\kappa_{\beta]}$ of SL(4,R) should be present in the dynamical holonomy group of a physical interaction connection.

If we designate the even subgroup, generated by $\kappa_{[\alpha}\kappa_{\beta]}$, by *L*, the previous discussion means that

$$L\subseteq H\subseteq G \quad . \tag{2.4}$$

Furthermore, *H* must be simple. The different possibilities may be obtained from the knowledge of the subgroups of *G*.

The possible simple subgroups are as follows:

1. The 10 dimensional group *P* generated by $\kappa_\alpha$, $\kappa_{[\beta}\kappa_{\gamma]}$. This group is isomorphic to the groups generated by, $\kappa_{[\alpha}\kappa_\beta\kappa_{\gamma]}$, $\kappa_{[\alpha}\kappa_{\beta]}$ and by $\kappa_{[a}\kappa_b\kappa_{c]}$, $\kappa_{[a}\kappa_{b]}$, $\kappa_5$ and in fact to any subgroup generated by a linear combination of these three generators;
2. The 6 dimensional subgroup *L*, corresponding to the even generators of the algebra, $\kappa_{[\alpha}\kappa_{\beta]}$. This group is isomorphic to the subgroups generated by $\kappa_a$, $\kappa_{[a}\kappa_{b]}$, and by $\kappa_0\kappa_{[a}\kappa_{b]}$, $\kappa_{[a}\kappa_{b]}$ and in fact to any subgroup generated by a linear combination of these three generators;
3. The group *G* itself.

The *P* subgroup is Sp(4,R), as may be verified by explicitly showing that the generators satisfy the simplectic requirement [18 ]. This group is known to be homomorphic to SO(3,2), a De Sitter group. The *L* subgroup is Sp(2,C), isomorphic to SL(2,C). In addition there are only two simple compact subgroups of *G*, non dynamical, generated by $\kappa_{[a}\kappa_{b]}$ and $\kappa_0$, $\kappa_5$, $\kappa_1\kappa_2\kappa_3$, apart from the unidimensional subgroups.

Then we have only three possible dynamical holonomy groups: *L*, *P*, or *G*. For each case we have an equivalence class of connections and a possible physical interaction within the theory. Other holonomy groups are not dynamical, in the sense that they do not produce a geometrical accelerating action on matter, as determined by an observer boost. This is the case of a U(1) holonomy subgroup which may represent electromagnetism but does not provide, by its direct action, a geometric dynamics (force) on charged matter. The dynamics requires a separate Lorenz force postulate.

Since $L\subset P\subset G$, this scheme allows us to classify interactions geometrically in increasing order of complexity. This group chain has a symmetry because there is no unique way of identifying $L\subset G$ and $L\subset P$. The coset $G/L$ represents how many equivalent $L$ subgroups are in $G$. There is also an equivalence relation $R$ among the non compact generators of $G$, all of them equivalent to a boost generator or space-time external symmetry. The subspace obtained as the quotient of $G/L$ by this relation is the internal symmetry group of $L\subset G$,

$$\frac{G/L}{R} = SU(2) \quad . \tag{2.5}$$

Similarly the coset $P/L$ gives, as the internal symmetry of $L\subset P$, the group

$$\frac{P/L}{R} = U(1) \quad . \tag{2.6}$$

The total internal symmetry of the chain $L\subset P\subset G$ is the product of the two groups SU(2)⊗U(1) which coincides with the symmetry group of the weak interactions. There is no geometrical reason to identify the structure group of the theory with the symmetry group because they are different geometrical concepts. At the end we shall come back to discuss the physical significance of this symmetry.

## 3. Subexcitations Corresponding to Subgroups.

If we restrict the group to either the $P$ or $L$ subgroups, the corresponding frame matrices (subframes) are elements of the subgroup. The total frame $f$, which is an element $a$ of the algebra, decomposes into even and odd parts in accordance with the general Clifford algebra decomposition,

$$a = a_+ + \kappa^0 a_- \quad . \tag{3.1}$$

From previous results, the equations of motion for the even $f_+$ and odd $f_-$ parts of $f$ are under certain restrictions,

$$\kappa^\mu \partial_\mu f_+ = \kappa^\mu \Gamma_{\mu-} f_- = m f_- \quad , \tag{3.2}$$

$$\kappa^\mu \partial_\mu f_- = \kappa^\mu \Gamma_{\mu-} f_+ = m f_+ \quad , \tag{3.3}$$

implying that a frame for a massive particle must have odd and even parts. In our case, if we set $f_-$ equal to zero we obtain also that $m$ is zero. Therefore, for an $L$-subframe we have multiplying by $\kappa^0$ ,

$$\bar{\sigma}^\mu \partial_\mu f_+ = 0 \quad , \tag{3.4}$$

the equation of the neutrino as discussed before.

If an excitation corresponds to a representation of a subgroup with specific quantum numbers, it may be associated to only one of the spinor columns of the frame, the one with the corresponding quantum numbers. Accordingly, we restrict the fluctuations of frames to matrices that have only one column in each of the two parts of the frame, the even $\eta$ and the odd $\xi$.

$$\eta = \begin{bmatrix} \eta_1^{\hat{1}} & 0 \\ \eta_1^{\hat{2}} & 0 \end{bmatrix} \quad , \tag{3.5}$$

$$\xi = \begin{bmatrix} \xi_1^{\hat{1}} & 0 \\ \xi_1^{\hat{2}} & 0 \end{bmatrix} \quad . \tag{3.6}$$

We now restrict to the even simple subgroup SL(2,C), homomorphic the Lorentz group. As shown in previ-ous work, the $\eta$, $\xi$ parts have inequivalent transformations under this group,

$$\eta' = l\eta \quad , \tag{3.7}$$

$$\xi' = l^*\xi \quad . \tag{3.8}$$

These columns are spinor representations of the group.

We may form a four dimensional (Dirac) spinor by adjoining the two spinors, where the components $\eta$, $\xi$ are two complex 2-spinors. We may combine the 2 columns into a single column Dirac 4-spinor,

$$\psi = \begin{bmatrix} \xi_1^{\hat{1}} \\ \xi_1^{\hat{2}} \\ \eta_1^{\hat{1}} \\ \eta_1^{\hat{2}} \end{bmatrix} \quad . \tag{3.9}$$

We now show that the even and odd parts of a frame are related to the left and right handed components of the field. We calculate the left handed and right handed components, and obtain, omitting the indices,

$$\psi_L = \tfrac{1}{2}\left(1+\gamma^5\right)\psi = \tfrac{1}{2}\left(1+\gamma^5\right)\begin{pmatrix}\xi \\ \eta\end{pmatrix} = \begin{pmatrix}0 \\ \eta\end{pmatrix} \quad , \tag{3.10}$$

$$\psi_R = \tfrac{1}{2}\left(1-\gamma^5\right)\psi = \tfrac{1}{2}\left(1-\gamma^5\right)\begin{pmatrix}\xi \\ \eta\end{pmatrix} = \begin{pmatrix}\xi \\ 0\end{pmatrix} \quad . \tag{3.11}$$

We have that the left handed component is equivalent to the $\eta$ field which in turn is defined in terms of the even $f_+$ field. Similarly, we see that the right handed component is equivalent to the $\xi$ field and consequently to the odd $f_-$ field. Therefore, an even frame corresponds to a left handed particle, as should be for a neutrino. The Lorentz frame excitation has neutrino like properties. It is necessary to point out that, since the mass in this equation is zero, Wigner's little group, which preserves the momentum $k$, is ISO(2) instead of SO(3) corresponding to massive particles. The induced representations of the neutrino excitation are sections over the light 3-hypercone valued in representations of ISO(2) and are characterized by the value of helicity.

A general Sp(4,R) excitation has a $\kappa^0$ generator, which is an electromagnetic generator. We expect that this excitation should represent a particle with one quantum of charge and one quantum of spin. The component of the current along this generator coincides with the standard electric current [19] in quantum mechanics. In addition, it was shown that the general perturbation technique for the interaction of the electron and neutrino fields lead, essentially, to the current and Lagrangian assumed in Fermi's theory of weak interactions [20] of leptons. This lead to the conjecture that a Sp(4,R) excitation may represent an electron. With this in mind, the value of the ratio of the proton to electron masses was derived from the definition of mass in terms of energy as indicated in a previous section, using the properties of the structure group and its subgroups and respective cosets in the theory.

It is possible to define the eigenvalue of the odd operator in the geometric algebra as a quantum number called oddity. Furthermore, transformations by the subgroup SL(2,C) leave invariant the even and odd subspaces of a frame. Since these eigenspaces correspond to the left handed and right handed parts, or equivalent to the helicity eigenvalues, it is clear that, within a Sp(4,R) frame, $e_L$ may be assigned a $\nu_L$ partner and may be considered a doublet, under another group, while $e_R$ has no $\nu_R$ partner and may be considered a singlet. This indicates a possible link with the standard model of weak interactions.

# 4. The General Material Excitation.

In general, the equation of motion for matter,

$$2\kappa^\mu \nabla_\mu e + \kappa^{\hat{\alpha}} \nabla_\mu u_{\hat{\alpha}}^\mu e = 0 \quad , \tag{4.1}$$

applies to the three classes of dynamical frames, according to the three dynamical holonomy groups. The three equations, together with the non linear field equation should have a substratum state solution. Then we may associate an excitation to each class of frames around the substratum solution. This excitations are elements of the Lie algebra of the structure group of the theory.

The maximally commuting subspace of the Lie algebra sl(4,R), generated by the chosen regular element, [21] is a three dimensional Cartan subalgebra, which is spanned by the generators,

$$X_1 = \kappa_1 \kappa_2 \quad , \tag{4.2}$$

$$X_2 = \kappa_0 \kappa_1 \kappa_2 \kappa_3 \quad , \tag{4.3}$$

$$X_3 = \kappa_0 \kappa_3 \quad . \tag{4.4}$$

It is clear that $X_1$, and $X_2$ are compact generators and therefore have imaginary eigenvalues. Because of the way they were constructed, they should be associated, respectively, to the $z$-component of angular momentum and the electric charge. Both of them may be diagonalized simultaneously in terms of their imaginary eigenvalues.

As shown in figure 1 the 4 members of the fundamental representation form a tetrahedron in the three dimensional Cartan space. They represent the combination of the two spin states and the two charge states of the associated particle. For example,

| charge | spin | flux | | |
|---|---|---|---|---|
| $-1$ | $+1$ | $-1$ | negative charge with spin up | $f_-^{\uparrow}$ |
| $-1$ | $-1$ | $+1$ | negative charge with spin down | $f_-^{\downarrow}$ |
| $+1$ | $+1$ | $+1$ | positive charge with spin up | $f_+^{\uparrow}$ |
| $+1$ | $-1$ | $-1$ | positive charge with spin down | $f_+^{\downarrow}$ |

The fundamental representation $f$ of SL(4,R), which may be indicated by $f_+^{\uparrow}$, $f_+^{\downarrow}$, $f_-^{\uparrow}$, $f_-^{\downarrow}$ groups together two excitation states of positive charge with two excitation states of negative charge. The presence of opposite charges in a representation forces us to make a clarification. To avoid confusion we should restrict the term (charge) *conjugation* to indicate the original Dirac operation to relate states of opposite charge.

There is also another fundamental representation $\tilde{f}$ dual to $f$ and of the same dimensions, with all signs reversed, which is inequivalent to the original one and represented by the inverted tetrahedron in the Cartan subspace. One of the two representations is arbitrarily assigned to represent the physical excitation. The mathematics of the representation algebra indicates that a state may also be described as composed of 3 members of its dual representation. In turn, the dual representation may also be similarly described as composed of 3 members of its dual, which is the original representation. This mutual mathematical decomposition is reminiscent of the idea of "nuclear democracy" proposed in the 1960's [22] but restricted to dual representations. To avoid confusion we restrict the use of the term *duality* to relate these inequivalent representations.

In standard particle language the antiparticle is the dual particle which implies also the conjugate particle. Nevertheless, this assignment of a physical particle/antiparticle pair to the fundamental representation and its inequivalent dual is not a mathematical implication of standard particle theory, only a physical assumption. Since our fundamental representation includes opposite charges it is not appropriate to consider the dual representation as antiparticles. We may, as well, simply say that the antiparticle is related to the conjugate particle. The dual particle is just a necessary dual mathematical structure.

We have then for a particular state $p$ of the fundamental representation,

$$p^{\uparrow} \equiv f_+^{\uparrow} = \left( \tilde{f}_+^{\uparrow} \quad \tilde{f}_+^{\downarrow} \quad \tilde{f}_-^{\uparrow} \right) \equiv \left( q_+^{\uparrow} \quad q_+^{\downarrow} \quad q_-^{\uparrow} \right) \quad , \tag{4.5}$$

in terms of the dual states $q$. Similarly for a particular $q$ state,

$$q_+^{\uparrow} = \left( \tilde{q}_+^{\uparrow} \quad \tilde{q}_+^{\downarrow} \quad \tilde{q}_-^{\downarrow} \right) = \left( p_+^{\uparrow} \quad p_+^{\uparrow} \quad p_-^{\downarrow} \right) \quad . \tag{4.6}$$

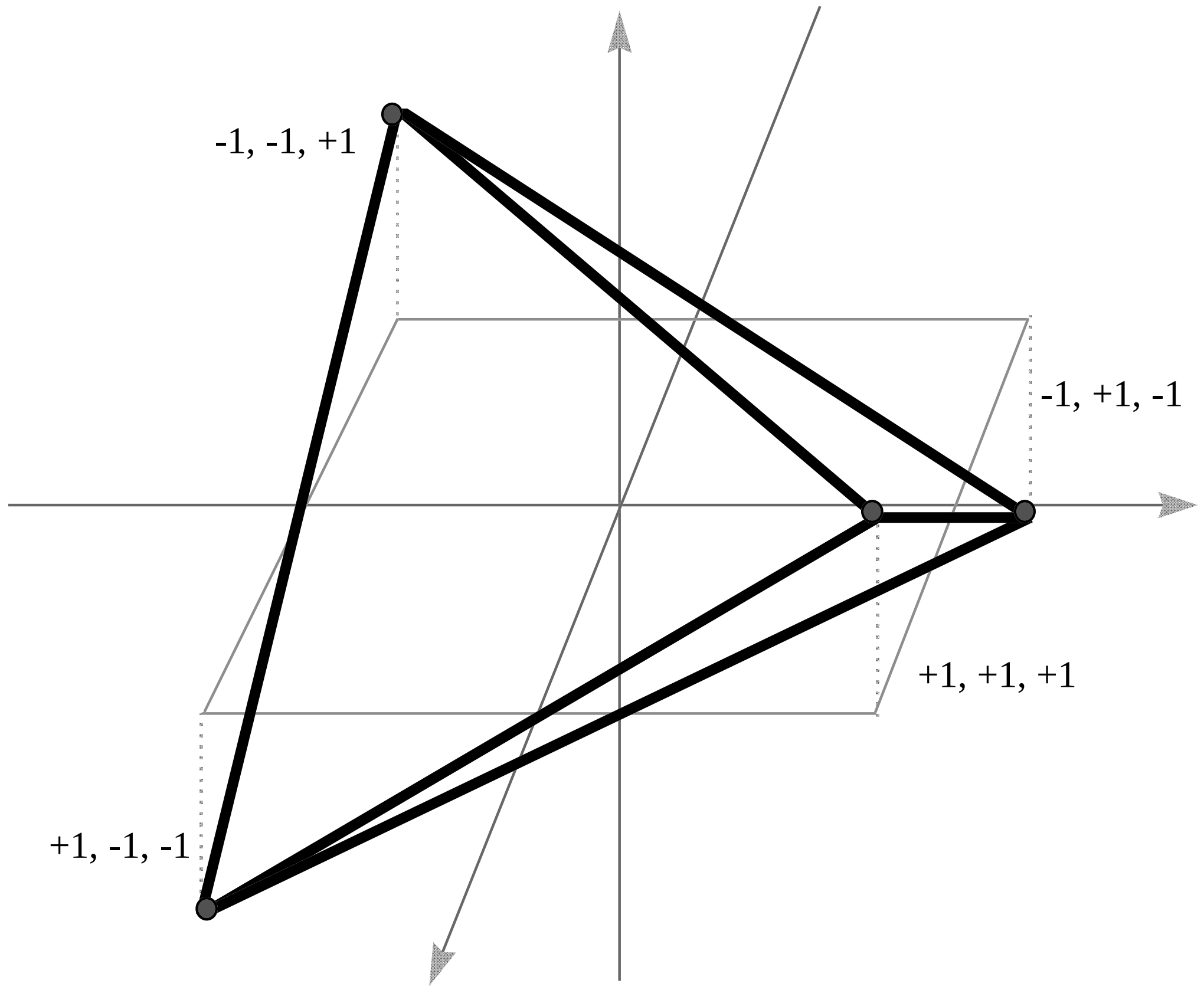

Figure 1

It does not follow that the *q* necessarily are states of a different physical excitation, only that the *q* form a dual triplet mathematical representation of the *p* representing the *same* excitation. This allows a different physical interpretation for these mathematical constructions. It should be noted that all *p*, *q* have electric charge equal to the geometric unit, electron charge $\pm e$. Since these excitations have particle properties, there is a dual mathematical representation of the physical excitation (particle). We may raise the following question? What happens if we physically identify *p* with the proton, which mathematically may be expressed as *3 q*, interpreted as quarks? In our theory there is no need to assign fractional charges to quarks. In accordance with the "restricted nuclear democracy" we may assume that the quark states, in turn, may be mathematically expressed as 3 protons *p*, which may justify the large experimental mass of these states.

The fundamental representation of Sp(4,R) which may be displayed as a tetrahedron projected square in a two dimensional Cartan space is $e_-^{\uparrow}$, $e_-^{\downarrow}$, $e_+^{\uparrow}$, $e_+^{\downarrow}$. Its dual Sp(4,R) representation, obtained by reversing all signs is mathematically the same $e_+^{\downarrow}$, $e_+^{\uparrow}$ $e_-^{\downarrow}$, $e_-^{\uparrow}$. Similarly the SL(2,C) representation $\nu^{\uparrow}$, $\nu^{\downarrow}$ which may be displayed in a one dimensional Cartan space is mathematically its own SL(2,C) dual. Therefore, the *only one* of the three excitations with a mathematically inequivalent dual structure is *p*. Since the Sp(4,R) and SL(2,C) subgroups may be imbedded in SL(4,R), the corresponding Cartan spaces of Sp(4,R) and SL(2,C) may also be imbedded in the Cartan subspace of SL(4,R). The plane subspace $Q=-1$ has charges of one definite sign and is a representation of Sp(4,R). Another plane, $Q=1$, containing opposite charges is another representation of Sp(4,R) that completes the SL(4,R) Cartan subspace (anti-particle $e_+$). States $q_+^{\uparrow}$, $q_+^{\downarrow}$ may be considered as an Sp(4,R) antimatter representation different from $e_-^{\uparrow}$, $e_-^{\downarrow}$, which is present inside the proton implying that there is no need to look for antimatter elsewhere. The analogous relations between the mathematical $q_+$ and $e_-$ structures as Sp(4,R) representations may explain the parallelism of hadron and lepton families. Nevertheless, they are different because the *q* are acted upon by the complete *G* connection whereas the *e* are acted only by the *P* connection.

It may be seen that the electron state $e_-^{\downarrow}$ is complementary to $q_+^{\uparrow}$ $q_+^{\downarrow}$ $q_-^{\uparrow}$ occupying the quartet of states of a SL(4,R) representation, reflecting the neutrality of matter composed of *p*, and *e*. The fundamental (1, ±1, ±1) SL(4,R) states define a proton charge sign. Since matter is neutral the fundamental (-1, ±1, ±1) Sp(4,R) states, imbedded in the proton Cartan space as subspace *Q*=-1, define the electron charge sign, opposed to the proton sign.

The space SL(4,R)/SU(2)⊗SU(2) is a 9-dimension Riemannian space of the non compact type. There are 9 boost generators $B^m{}_a$. The rotation SU(2), indicated by *S*, contained in SO(4) acts on the *m* index and the electromagnetic SU(2), indicated by *Q*, contained in SO(4) acts on the *a* index,

$$Q_b^a B_a^n S_n^m = B_b'^m \quad . \qquad \textbf{( 4.7)}$$

Similarly the subspace Sp(4,R)/SU(2)⊗U(1) is a 6-dimensional Riemannian space of the non compact type. The complementary subspace within SL(4,R)/SU(2)⊗SU(2) is 3-dimensional. There is a triple infinity of these subspaces within the total space, reflecting the triple infinity of groups *P* in *G*, depending on the choice of an electromagnetic generator among the three possible ones in SU(2).

In Sp(4,R)/SU(2)⊗U(1), with a *fixed* electromagnetic generator, we have a vector in the odd sector, representing a momentum *k* associated to an excitation *e*. Since we have this situation in SL(4,R)/SU(2)⊗SU(2) for each generator in SU(2), we have, in effect, 3 momenta, $k_i$, that *may* characterize an excitation *p*. We must consider excitations characterized by 3 momenta, $k_i$, which may be interpreted as three subexcitations.

The 3 momenta, $k_i$, of the three subexcitations characterize the protonic frame excitation *p*, but neither those of *e* nor those of *ν*. Mathematically what we have is a system characterized by 3 momenta, $k_i$, that may be scattered into another system of 3 momenta, $k_j$'. Could this be interpreted as a 3-point particle? It should be expected that a scattering analysis of excitations must include three momenta in some $\delta$ functions that appear in the scattering results. Experimentally, at high energies, this should appear as a collision with a system of three point like particles. Again, this could be interpreted as three partons (quarks) inside the proton. In any case, it is clear that a point like proton excitation is not predicted by our theory.

# 5. Physical Interpretation in Terms of Particles and Relativistic interactions.

We have found properties of the geometric excitations which are particle like. We take the position that this is no coincidence but indicates a geometrical structure for physics. The source current *J* depends on a frame. To each holonomy group we may associate a class of frames thus giving three classes of matter.

As already discussed in section 3, the corresponding *L*-frame represents a zero mass, neutral, spin ½, left handed geometric excitation, which obeys eq. ( 3.4), and has the particle properties of the neutrino.

For the *P*-connection, the corresponding *P*-frame represents a massive, opposite charge -1, spin ½, geometric point like excitation which has the properties of the electron [23].

For the *G*-connection the corresponding *G*-frame should be a massive, charge 1, spin ½, 3-point geometric excitation, with a bare mass of 1836.12 times the electron mass which we conjecture is the proton.

In this manner, we have associated to each of the three holonomy groups, one of the only three known stable particles.

For the *L*-connection it is not difficult to recognize that the interaction is gravitation, from the discussion in previous chapters and the work of Carmelli [24]. Similarly, it also was shown in previous work that electromagnetism (without dynamics) is associated to one of the SU(2) generators and that the physical Fermi weak interaction is related to the odd sector of a *P*-connection.

We propose here that the *P/L* generators may be interpreted as an electroweak interaction and the *G/P* as strong nuclear interaction. Then the three dynamic holonomy classes of connections may correspond to three classes of relativistic interactions as follows:

1. The *L*-connection describes gravitational interaction;
2. The *P*-connection describes coupled gravitational and electroweak interactions;
3. The *G*-connection describes coupled gravitational, electroweak and strong interactions.

The $L$-frames obey equations that may be obtained from the general equations of motion when the frame $e$ has only the even part $e_+$. From the field equation it is seen that a $P$-frame generates a $P$-connection and that a $G$-frame generates a $G$-connection.

From this classification it follows, in agreement with the physical interactions that:

1. All (matter) frames self interact gravitationally;
2. $L$-frames self interact only gravitationally (uncharged matter);
3. $G$-frames (hadrons) are the only frames that self interact strongly (hadronic matter);
4. $G$-frames self interact through all three interactions;
5. $G$-frames (hadronic matter) and $P$-frames (leptonic matter) self interact electroweakly and gravitationally;
6. $P$-frames self interact gravitationallly an electroweakly but not strongly (leptonic matter).

# 6. Particle Families.

The framework of this theory is compatible with a phenomenological classification of particles in a manner similar to what is normally done with the standard symmetry groups.

First we should notice that the theory suggests naturally three stable ground particles ($\nu$, $e$, $p$). In fact, if we consider the possibility of different levels of excited states, each particle may generate a class of unstable particles or resonances.

In particular, since the equations for each of the three particle classes are the same, differing only in the subgroup that applies, it may be expected that there is some relation among corresponding levels of excitations for each class, forming families.

The ratio of the mass of the hadron in the ground level family (proton) to the mass of the lepton in the ground level family (electron) has been calculated from the ratio of volumes of coset spaces based on the existence of a trivial non zero constant solution for the substratum connection. This trivial solution is the representative of a class of equivalent solutions generated by the action of the structure group.

Now consider only topological properties, independent of the connection, of the space of complete solutions (substratum plus excitation solutions). An incoming scattering solution is a jet bundle local section, over a world tube in the space time base manifold, that describes the evolution of the solution in terms a time like parameter $\tau$ from past infinity to some finite time $t$. Similarly, an outgoing solution is a local section from time $t$ to future infinity. The local sections in the bundle represent classes of solutions relative to local observers. Scattering solutions at infinity are asymptotically free excitation solutions around a substratum. The substrata (incoming and outgoing) are equivalent to each other and to the constant trivial solution if we choose observer frames adapted to the substrata.

Since the equations are of hyperbolic type, we should provide initial conditions on an initial 3 dimensional hypersurface at past infinity $\mathscr{I}_-$. The sections of interest are the three fundamental excitations. We require that all incoming solutions, at the past infinity hypersurface $\mathscr{I}_-$, reduce to a free excitation around the trivial substratum solution at the two dimensional spatial infinity subspace $\mathscr{I}_-(\infty)$. Since the incoming solution substrata are equivalent to the trivial substratum solution at spatial infinity $\mathscr{I}_-(\infty)$, we may treat this spatial infinity as a single point, thus realizing a single point compactification of $\mathscr{I}_-$, so the initial hypersurface $\mathscr{I}_-$ is homeomorphic to $S^3$. All incoming solutions on $\mathscr{I}_-$ are classified by the functions over $S^3$. The same requirements may be applied to the outgoing remote future solutions and, in fact, to any solution along an intermediate 3 dimensional hypersurface, a section of the world tube. Thus, the final hypersurface at future infinity $\mathscr{I}_+$ is also homeomorphic to $S^3$. The incoming and outgoing substratum local sections over $\mathscr{I}_-$ and $\mathscr{I}_+$ must be pasted together in some common region around the present $t$, by the transition functions of the bundle. All generators of the group produce a transformation to a different, but equivalent under the group, expression for the solution. If the holonomy group of the solution is not the whole group, there is a reference frame that reduces the structure group to the particular holonomy subgroup. But in general for arbitrary observer, there are solutions formally generated by SL(4,R). Since this transition region, the "equator", has the topology of $S^3 \times R$, the transition functions $\varphi$ define a mapping, at the $\tau=0$ hypersurface,

$$\varphi : S^3 \to SL(4,R) \quad , \tag{6.1}$$

which is classified by the third homotopy group [25 ] of the structure group SL(4,R) or the respective holonomy subgroup. There are some solutions not deformable to the trivial solution by a homeomorphism because $\varphi$ represents the twisting of local pieces of the bundle when glued together.

To determine $\pi_3(G)$, we realize that the exponential map from the maximally non compact subalgebra of sl(4,R) is a diffeomorphism [26] to the non compact Riemannian coset subspace which is contractile. We have a short exact sequence in the general homotopy sequence [27],

$$\cdots\pi_4\left(G/H\right)\xrightarrow{\Delta_*}\pi_3(H)\xrightarrow{i_*}\pi_3(G)\xrightarrow{p_*}\pi_3\left(G/H\right)\ldots\quad , \tag{6.2}$$

$$\{0\}\longrightarrow\pi_3(H)\xrightarrow{*}\pi_3(G)\longrightarrow\{0\}, \tag{6.3}$$

which implies that the intermediate mapping is an isomorphism and

$$\pi_3(G)=\pi_3(H) \tag{6.4}$$

where $H$ is the maximal compact subgroup. We also know that there is an isomorphism between the homotopy groups of a group and its covering group, except for the first homotopy group [28]. For the homotopy group of SL(4,R) we get

$$\pi_3(SL(4,R))=\pi_3(SU(2)\otimes SU(2))=\pi_3(SU(2))\otimes\pi_3(SU(2))=Z\otimes Z\,. \tag{6.5}$$

Similarly, we have for the homotopy groups of the other two holonomy groups,

$$\pi_3(Sp(4,R))=\pi_3(R\otimes SU(2))=\pi_3(SU(2))=Z\,, \tag{6.6}$$

$$\pi_3(SL(2,C))=\pi_3(SU(2))=Z\,. \tag{6.7}$$

These scattering solutions are characterized by topological quantum integral numbers $n$, for the three groups, and $m$ only for group $G$, called winding numbers. In all cases the scattering solutions are characterized by one topological winding number $n$ and in particular the hadronic scattering solutions have an additional topological winding number $m$. This result implies that there are solutions $\omega_n$, $e_n$ that are not homotopically equivalent to $\omega_0$, $e_0$.

All $p$, $e$, $\nu$ excitation solutions with a given $n$ may be associated among themselves because of the isomorphism of the homotopy groups $Z$. This is an equivalence relation. Two $p$, $e$, $\nu$ excitation with the same $n$ are in the same topological class determined by the substratum. Each topological class, characterized by the topological quantum number $n$ defines a physical class, a family of particles with the same winding number, which is respected by transitions the same as the algebraic quantum numbers $s$, $q$, $f$.

For example, the association of a solution of each type to any value of $n$ may be part of a general scheme of relations labeled by $n$ as follows,

$$\begin{array}{llll} l_0=e, & l_1=\mu, & l_2=\tau, & \cdots \\ \nu_0=\nu_e & \nu_1=\nu_\mu & \nu_2=\nu_l & \cdots \\ h_0=p_e & h_1=p_\mu & h_2 & \cdots \end{array}, \tag{6.8}$$

among the excitation levels of the electron (proper leptons), the excitation levels of the neutrinos (other neutrinos) and the excitation levels of the proton (hadrons).

# 7. Relation with Barut's Model.

Since in our geometric theory the only constant that enters is $\alpha$, the ratio of the masses of the excited states and the fundamental state should depend on this constant. It is possible that the geometric theory may provide an explanation for the interesting approximate relation of the masses of the leptons and the electron,

$$m_l=m_e\left(1+\frac{3}{2\alpha}\sum_0^l n^4\right)\quad . \tag{7.9}$$

This equation was proposed by Barut [29] with $n$ interpreted as the number of $\bar{\nu}\nu$ pairs. Although this interpretation is also characterized by the group Z, it is not clear whether both interpretations are compatible.

In any case, the possibility exists that there be excited states which may be interpreted as a particle composed of $p$, $e$ and $\nu$ . As a matter of fact Barut has suggested that the muon should be considered as an electromagnetic excitation of the electron. Although the details should be different, because Barut's approach is only electromagnetic in nature and ours is a unified interaction, we may conjecture also that the muon is an excited electron state with a leptonic winding number *n=1.* Similarly the $\tau$ would be an excitation with winding number *n=2.*

Having a possible scheme for the geometric origin of the muon, it is possible that the process of building other particles as suggested by Barut [30 ,31 ,32 ] may be accomplished in terms of the stable articles, the electron, the proton and the neutrino together with the muon.

In order to adapt our geometric model to Barut's model we establish that the quantum numbers of an excitation correspond to the quantum numbers of its stable components ($p$, $e$, $\nu$). In this manner, following Barut's ideas, we define the charge of an excitation as the number of protons minus the number of electrons contained in the excitation,

$$Q = N_p - N_e \quad . \qquad (7.10)$$

Similarly, the barionic number of an excitation is the number of protons in the excitation,

$$B \equiv N_p \qquad (7.11)$$

and the leptonic number of an excitation is the number of electrons and neutrino components,

$$L \equiv N_e + N_\nu \quad . \qquad (7.12)$$

Strangeness of an excitation is the number of excited electrons (winding number 1), or muons $\mu$ in the excitation,

$$-S \equiv N_\mu \qquad (7.13)$$

and charm is the number of muonic neutrinos,

$$C = N_{\nu(\mu)} \quad . \qquad (7.14)$$

The isotopic spin depends on the number of stable particles in the following manner,

$$2I_3 \equiv N_p - N_e + N_\nu \quad . \qquad (7.15)$$

From these definitions it is possible to derive the Gell-Mann Nishijima formula,

$$Q = I_3 + \tfrac{1}{2}(B + S) \quad . \qquad (7.16)$$

and define strong hypercharge

$$Y \equiv B + S \quad . \qquad (7.17)$$

In additon to these Barut definitions, it is also possible to define weak hypercharge for a $P$-excitation in terms of leptonic number and the oddity $O$ of the representative algebraic element,

$$y \equiv -(L + O) \quad . \qquad (7.18)$$

The hadronic states contain protons. In figure 2 we show the barionic octet $J^P = {1/2}^+$ and in figure 4 the barionic decuplet $J^P = {3/2}^+$ . The meson states are bound states of two leptons $l\bar{l}$ as indicated in the mesonic octect $J^P = 0^+$ in Figure 3.

Furthermore, the geometric theory allows a discussion of the approximate quantum interaction between two of quark excitations forming a two 3-velocity system (Quarkonium?). We speculate that the non relativistic effective potential should be similar to the one in QCD because there are similarities of the mathematics of the theories. If this were the case, (this has to be shown) the non relativistic effective potential would be [33 ]

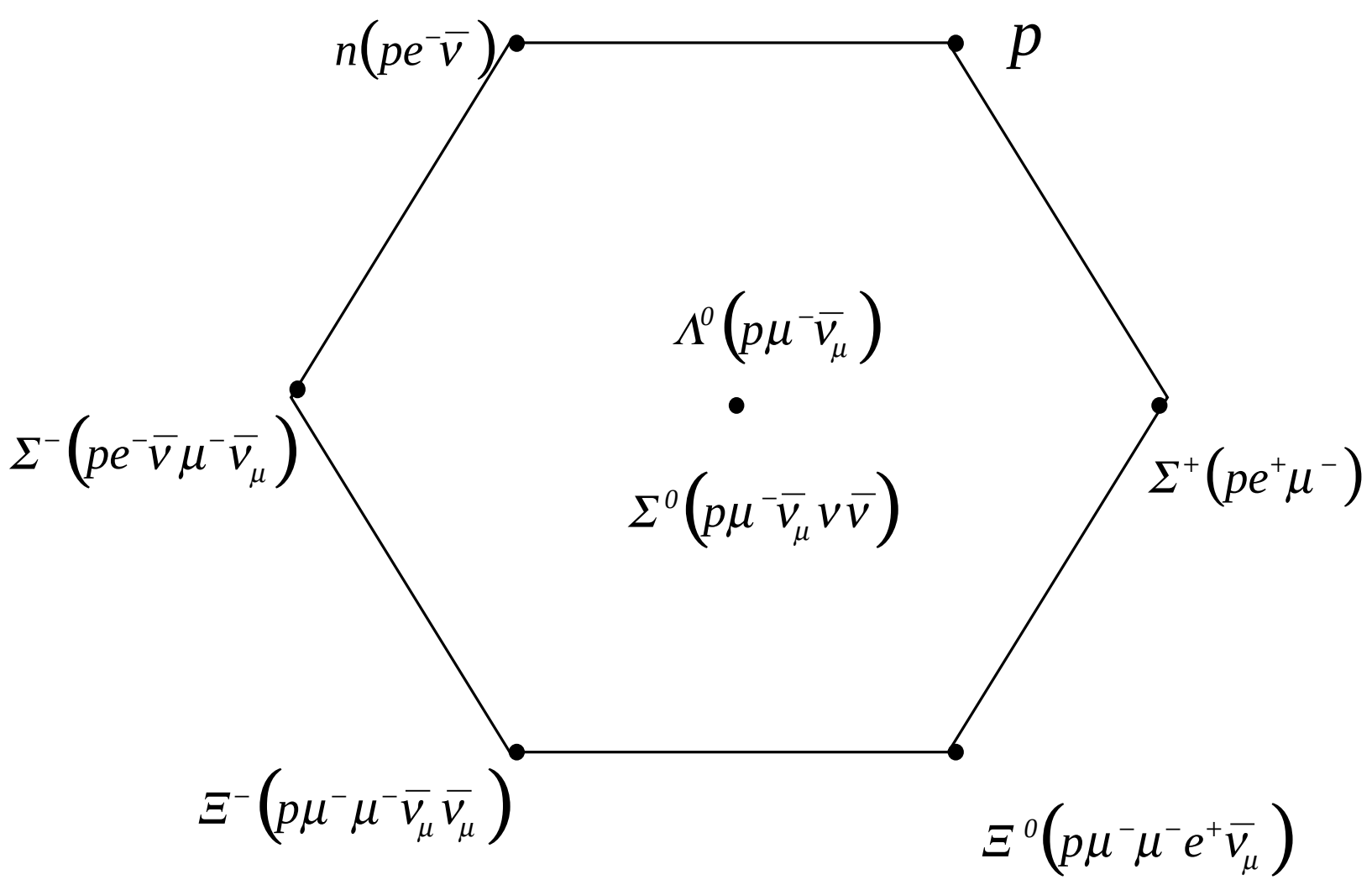

Barionic octet

Figure 2

$$(r) = -\frac{a}{r} + kr \qquad \textbf{( 7.19)}$$

and then we may fit the experimental $\Psi$ excited levels accordingly as is done in QCD.

The full frame excitation has proton like properties. Of course, within the geometric theory the quark excitations are not the fundamental building blocks of matter. They are only approximate subexcitations that are useful and necessary in describing a series of hadronic excitations.

## 8. Relation with the Standard Model.

We have replaced the space time related structure group, fundamental to the theory, by using SL(4,R) instead of SO(3,1). On top of this geometry, as is done in the standard model over the geometry of standard special relativity, it is possible to add a related structure to help in the understanding of the physical particles. It is known that heavy nuclei may be partially understood by using groups SU(N), U(N), O(N) to associate protons and neutrons in a dynamic symmetric or supersymmetric manner [34]. This is essentially the use of group theory to study the combinations of the two building blocks, protons and neutrons, assumed to form nuclei. In the same manner we *may* use certain groups to describe the combinations of the three fundamental geometrical building blocks introduced by the physical geometry, protons, electrons and neutrinos, assuming they form other particles.

We are interested in how to combine $G$, $P$ and $L$ excitations. This is accomplished by attaching $L$ excita-

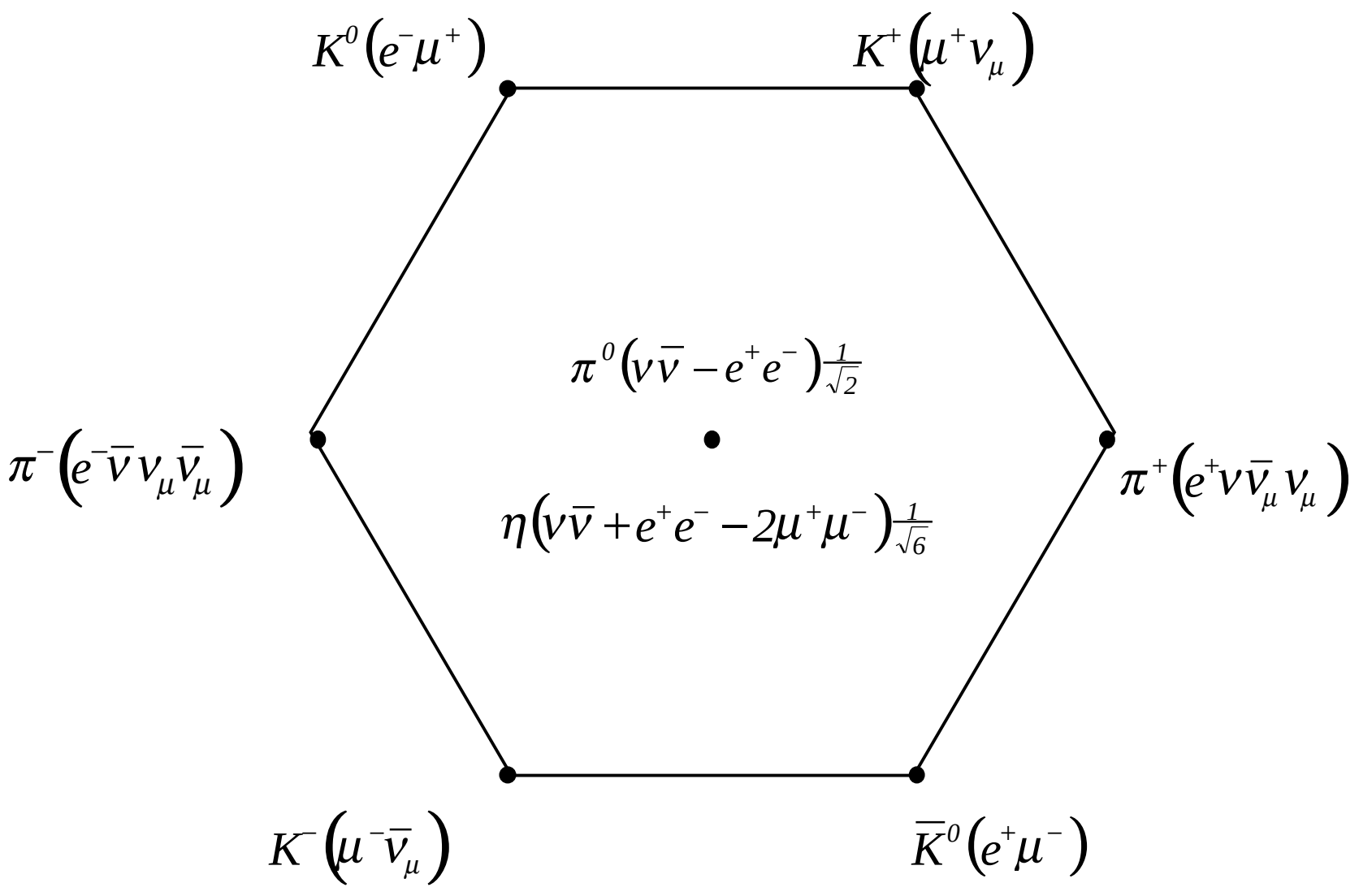

Pseudo scalar mesonic octet

Figure 3

tions to $P$ excitations and then to $G$ excitations. This combination may not be done uniquely, rather it depends on the identification of a subgroup $H$ with a subspace in the group fiber bundle space $G$. Any non compact generator is equivalent to a boost, by the adjoint action of a compact generator. Thus, the compact generators generate a symmetry of the combination of excitations.

In particular, the identification of $L$ within $P$ is not unique. The group $L$ may be expressed as the principal bundle $(L, {}^3B, S)$, where the fiber $S$ is the $SU(2)_R$ associated to rotations and ${}^3B$ is the three dimensional boost symmetric space,

$$ {}^3B = \frac{SL(2,C)}{SU(2)_R} \qquad (8.1) $$

The group $P$ may be expressed as the principal bundle $(P, {}^6B, S\otimes U(1))$ where ${}^6B$ is the six dimensional double boost symmetric space,

$$ {}^6B = \frac{Sp(4,R)}{SU(2)_R \otimes U(1)_Q} \qquad (8.2) $$

The choice of the boost sector of $L$ (even generators), inside $P$, depends on the action of the U(1) group in the fiber of $P$, generated by $\kappa_0$, a rotation within the double boost sector in $P$.

$$ [\kappa_0, \kappa_1] = 2\kappa_0\kappa_1 \quad . \qquad (8.3) $$

Mathematically this corresponds to the adjoint action of the only compact generator in the odd quotient $P/L$.

Similarly, the identification of $P$ inside $G$ is not unique either. The group $G$ may be expressed as the bundle

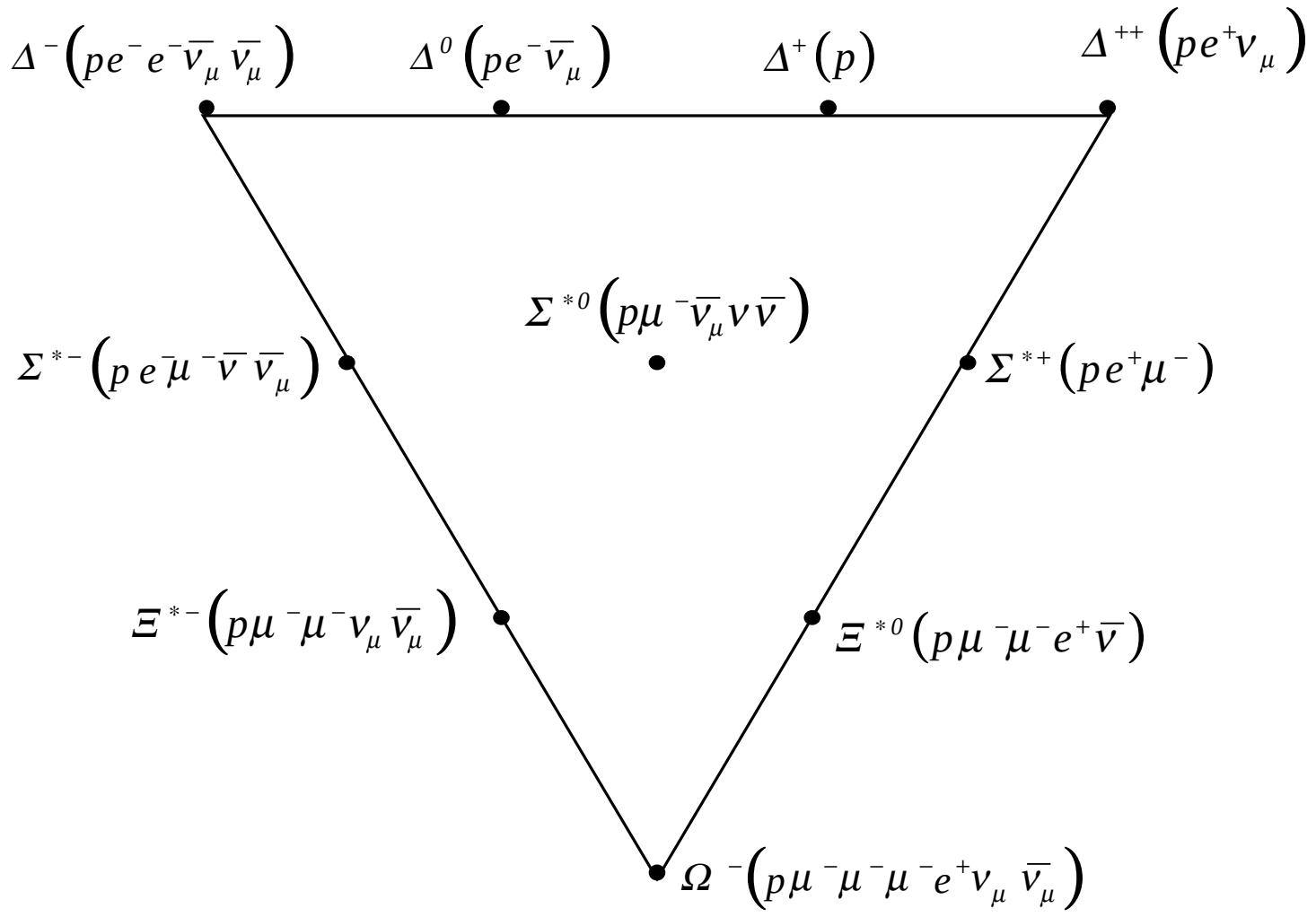

Barionic decuplet

Figure 4

($G$,${}^9B$,$S\otimes Q$) where $Q$ is the $SU(2)_Q$ associated to charge and ${}^9B$ is the nine dimensional triple boost symmetric space,

$$ {}^9B = \frac{SL(4,R)}{SU(2)_R \otimes SU(2)_Q} \qquad \textbf{(8.4)} $$

The choice of the boost sectors of $L$ and $P$ sector inside $G$ depends on the action of the SU(2) in the fiber of $G$, generated by the compact electromagnetic generators. Mathematically this corresponds to the action of all compact generators in the quotient $G/L$. The total symmetry of first identifying $L\supset P$ and then $(L\supset P)\supset G$ is the product of the two groups. Therefore, there is a symmetry action of a group equal to SU(2)⊗U(1), *but different to the subgroups of G*, on the identification of the subgroups in the chain $L\supset P\supset G$.

There is an induced symmetry on the combination of $L$ and $P$ subexcitations on $G$ excitations corresponding to this SU(2)⊗U(1) symmetry. The action of this combinatorial group may be interpreted as the determination of the possible combinations of $L$ and $P$ excitations to give flavor to states of $G$ and $P$ excitations and thus related to the weak interactions.

This $SU(2)_F$ group relates electron equivalent $P$-excitation states or neutrino equivalent $L$-excitation states. Clearly at sufficiently high energies the mass of any excitation kinematically appears very small and its effects on results are small deviations from those of a zero mass excitation that always corresponds to an excitation of the even subgroup or $L$-excitation. For this reason, at high energies, the even part (left handed part) of a $P$-excitation may be related to an even $L$-excitation by an SU(2) transformation. Both corresponding physical leptonic excitations, the even part $e_+$ or $e_0$ and $\nu$, may be considered members of a doublet, labeled by oddity 0 or weak hypercharge -1, while the odd part $e_-$ or $e_1$ is a singlet labeled by oddity 1 or weak hypercharge -2. This weak interaction association of leptons into hypercharged states has the approximate symmetry $SU(2)_F$. In addition, there is an undetermined orientation of $L_1$ inside $G$ that depends of the action of SU(2)⊗U(1). Under this approach, the standard physical electromagnetic potential $A$, the connection of $U_+(1)$ in $SL_1(2,C)$, has an orientation angle within the chiral SU(2)⊗U(1).

Since our description may be made in terms of the *proton representation or, alternately, in terms of its dual quark representation* as indicated in section 4, we could obtain complementary dual realizations of the relations among different experimental results. These are really different perceptions or pictures of the same physical reality of matter. The proton or $G$-excitation has a quark structure: it behaves as if formed by three points. Using the quark representation to build other particles, we may consider that in $G$ the 3 quark states span a 2 dimensional Cartan $A_2$ subspace. There is a combinatorial unitary symmetry, SU(3), related to this Cartan subspace, that may be interpreted as the color of the combinations of states of $G$ and thus related to strong interactions. States of $G$-excitations display an $SU(3)_C$ symmetry.

In this manner, the physical geometry determines an approximate combinatorial internal symmetry characterized by the combinatorial symmetry group $S_C$.

$$S_C = SU(3)_C \otimes SU(2)_F \otimes U(1) \quad . \qquad \textbf{(8.5)}$$

If the probability of combination is propagated infinitesimally by a connection, we have the elements of a high energy quasi standard model. At high energies we may expect the appearance of resonances with high masses. Of course, in order to get the electromagnetic potential at low energy, the chiral part of the phenomenological combinatorial internal symmetry should be broken.

The standard model is built from certain general features of experimental and theoretical results. In particular the SU(3) transformation among a triplet of components, the quark-parton model [35 ,36 ], and the $SU(2)_L$ transformation of the chiral parts of a doublet, the electroweak theory [37 ,38 ], are the starting points of the model. Since these features are present in the physical geometry, as shown here and in sections 3 and 4, we have the real possibility of building an approximate "quasi standard model" on top of our geometry. The combinatorial groups actually relate discrete asymptotic, in or out, states in a scattering theory of excitations. It is possible, in an approximate way, to "gauge" these groups to obtain an approximate dynamic theory. Theoretically, the infinitesimal evolution of physical systems or dynamics is determined by the generators of SL(4,R) in accordance with the equations of the theory, but the scattering solutions should show symmetries related to the combinatorial group.

In this manner, this approximate construction should share certain features and results with the standard model while there would be differences in some other features. Due to the vastness of particle physics experimental results, it is not clear, at this time, how these differences would compare with experiments, in particular because the geometrical ideas may introduce a rearrangement of experimental results. This actually represents a program to be carried out. It is true that the standard model has led to success, but this does not imply there is no better way of reordering the physical results.

## Summary.

The geometric unified theory discussed represents relativistic interactions that may be classified into three classes by the dynamical holonomy groups of the possible physical connections. The three classes correspond to gravitation, electro-weak and strong physical interactions.

Solutions to the three corresponding frame excitation equations representing matter, show algebraic and topological quantum numbers. Using them and previously calculated mass ratios, we are able to identify three classes of fundamental excitations, corresponding to the three stable physical particles (neutrino, electron and proton). The algebraic numbers correspond to electric charge, spin and magnetic flux. The leptonic and hadronic topological quantum winding numbers correspond to higher levels of excitation, defining families for each integer. The interactions felt by these excitations conform with the general classification scheme of particles into neutrinos, leptons and hadrons. The combinations of the three particles display an SU(3)⊗SU(2)⊗U(1) symmetry. A program to establish the relation of the theory with the standard model of particles was outlined.